# Near-Sun Speed of CMEs and the Magnetic Non-potentiality of their Source Active Regions

Sanjiv K. Tiwari[1], David A. Falconer[1,2], Ronald L. Moore[1,2], P. Venkatakrishnan[3], Amy R. Winebarger[1], and Igor G. Khazanov[2]

We show that the speed of the fastest coronal mass ejections (CMEs) that an active region (AR) can produce can be predicted from a vector magnetogram of the AR. This is shown by logarithmic plots of CME speed (from the SOHO LASCO CME catalog) versus each of ten AR-integrated magnetic parameters (AR magnetic flux, three different AR magnetic-twist parameters, and six AR free-magnetic-energy proxies) measured from the vertical and horizontal field components of vector magnetograms (from the *Solar Dynamics Observatory's Helioseismic and Magnetic Imager*) of the source ARs of 189 CMEs. These plots show: (1) the speed of the fastest CMEs that an AR can produce increases with each of these whole-AR magnetic parameters, and (2) that one of the AR magnetic-twist parameters and the corresponding free-magnetic-energy proxy each determine the CME-speed upper-limit line somewhat better than any of the other eight whole-AR magnetic parameters.

## 1. Introduction

One of the most challenging tasks in the field of modern space research is the prediction of the severity of geomagnetic storms and solar energetic particle storms caused by solar flares and coronal mass ejections (CMEs). Active regions (ARs) on the Sun are the main sources of the biggest flares and most energetic CMEs [*Zirin and Liggett*, 1987; *Subramanian and Dere*, 2001; *Falconer et al.*, 2002; *Venkatakrishnan and Ravindra*, 2003; *Guo et al.*, 2007; *Wang and Zhang*, 2008; *Gopalswamy et al.*, 2010]. The initial speed of CMEs is one of the most important parameters that (among others e.g., the direction, width and mass of CMEs, orientation and strength of magnetic field therein) can help forecast the severity of geomagnetic storms and particle storms [see e.g., *Srivastava and Venkatakrishnan*, 2002; *Gopalswamy et al.*, 2010; *Dumbović et al.*, 2015, and references therein]. The magnetic non-potentiality of an AR, inferred by, for instance, free magnetic-energy proxies and magnetic-twist parameters, is most likely to determine the initial speed of CMEs emanating from the AR. Several other unexplored parameters e.g., AR lifetime, flux emergence/cancellation [e.g., *Subramanian and Dere*, 2001] might be important as well. Therefore, study of the relationship between properties of the photospheric magnetic field of an AR and the physical properties of the CMEs produced by the AR, e.g., their initial speed, is of great importance for forecasting severe space weather.

*Venkatakrishnan and Ravindra* [2003] estimated the potential-magnetic-field energy of 37 ARs from their line-of-sight (LOS) magnetograms and found it to be a reasonable predictor of the speed of CMEs arising from the ARs. The present paper reports a similar but more extensive investigation based on vector magnetograms instead of LOS magnetograms. *Liu* [2007] studied 21 halo CMEs and found a positive correlation of free magnetic energy of ARs with CME speed. CME speed is also found to be correlated with the GOES X-ray magnitude of the co-produced flare [*Ravindra*, 2004; *Burkepile et al.*, 2004; *Vršnak et al.*, 2005; *Gopalswamy et al.*, 2007; *Bein et al.*, 2012]. *Tiwari et al.* [2010] found a good correlation between a twist parameter [spatially averaged signed shear angle: *Tiwari et al.*, 2009a] of ARs and the GOES X-ray magnitude of flares produced by the ARs. A comparison of results from *Tiwari et al.* [2010] and *Jing et al.* [2010] suggests that this global twist parameter is strongly correlated with the free magnetic energy of ARs. From the above, one expects twist parameters and free-energy proxies to be determinants of the speed of the CMEs from an AR, and therefore determinants of the severity of the resultant geomagnetic storms, based on the results of *Srivastava and Venkatakrishnan* [2002]. In the present analysis, we investigate the relationship between magnetic parameters of ARs (mainly various twist parameters and free-energy proxies) and the initial speed of CMEs arising from the ARs.

Major CMEs emanating from ARs are co-produced with a flare [*Yashiro et al.*, 2008; *Wang and Zhang*, 2008; *Schrijver*, 2009]. Although several investigations have focused on predicting the flares from ARs by measuring various magnetic non-potentiality parameters [see e.g., *Hagyard et al.*, 1984; *Canfield et al.*, 1999; *Falconer et al.*, 2002; *Georgoulis and Rust*, 2007; *Leka and Barnes*, 2007; *Falconer et al.*, 2009; *Moore et al.*, 2012; *Falconer et al.*, 2014; *Bobra et al.*, 2014; *Bobra and Couvidat*, 2015], a direct link of any of these magnetic parameters to CME parameters have not been established thus far. To establish such a relationship requires 1. a careful manual inspection of which CME comes from which AR, 2. analysis of vector magnetograms of source ARs within 45 heliocentric degrees of disk center. In the present work, we first generated a list of a large number of CMEs that were observed by the SOHO LASCO/C2 coronagraph and were identified with flares in ARs observed by the Solar Dynamics Observatory (SDO). We then manually inspected those CMEs to find the CMEs that came from a clearly identified source AR or sometimes two neighboring ARs. We then calculated different twist parameters and free-energy proxies using vector magnetograms from SDO's Helioseismic and Magnetic Imager [HMI: *Schou et al.*, 2012; *Hoeksema et al.*, 2014], and studied their relationships to initial CME speeds collected from the LASCO/CME catalog [*Gopalswamy et al.*, 2009].

[1]NASA Marshall Space Flight Center, ZP 13, Huntsville, AL 35812, USA.
[2]Center for Space Plasma and Aeronomic Research, University of Alabama in Huntsville, Huntsville, AL 35805, USA.
[3]Udaipur Solar Observatory, Physical Research Laboratory, Udaipur 313001, India.

## 2. Event Selection and Data Analysis

First, we determined from the online LASCO/CME catalog (http://cdaw.gsfc.nasa.gov/CME_list/) all CMEs that took place between the start of the SDO mission (May 2010) through March 2014 (as far as the LASCO/CME catalog covered at the time of our analysis). We identified all CMEs that had a plane-of-sky width greater than 30 degrees, and had a co-produced flare in an AR identified by NOAA. Further, the flaring AR had to be between 45E to 45W, and the flare occurred ($t_{flare}$) up to 2 hours before the recorded start time till half an hour after the recorded start time of the CME ($t_{cme}$) in images from the LASCO/C2 coronagraph ($t_{cme}-2\text{hrs} < t_{flare} < t_{cme}+ 30$ min). The broad window for automatic selection was chosen, so as to not accidentally eliminate a CME/flare combination before we manually checked it. We found 946 CMEs following our criteria during the given time period of observation by SDO. For each of the 946 automatically selected CMEs, we manually verified that: (1) the CME was not seen in the LASCO C2 before the flare took place, (2) the CME occurred in the same quadrant as the source AR, and (3) there was no second flare occurring in another AR at nearly the same time. If there was a second flaring AR, we further verified that it was not the source of the CME under investigation.

By looking at LASCO-C2 movies and GOES X-ray flux plots, we made sure that the prospective flaring source AR was present on the frontside of the Sun. By looking at STEREO A & B movies we ensured that the CME was directed towards Earth; it did not come from a source on the back of the Sun. Far-side CMEs were discarded. We then used AIA 193 Å movies to determine which AR flare (out of sometimes several listed) was co-produced with the CME under investigation. The selection procedure for an example CME is illustrated in Figure 1. The movies (both for STEREO A & B and AIA 193) for the example event, shown in Figure 1, can be found at: http://cdaw.gsfc.nasa.gov/CME_list/daily_movies/2012/07/12/.

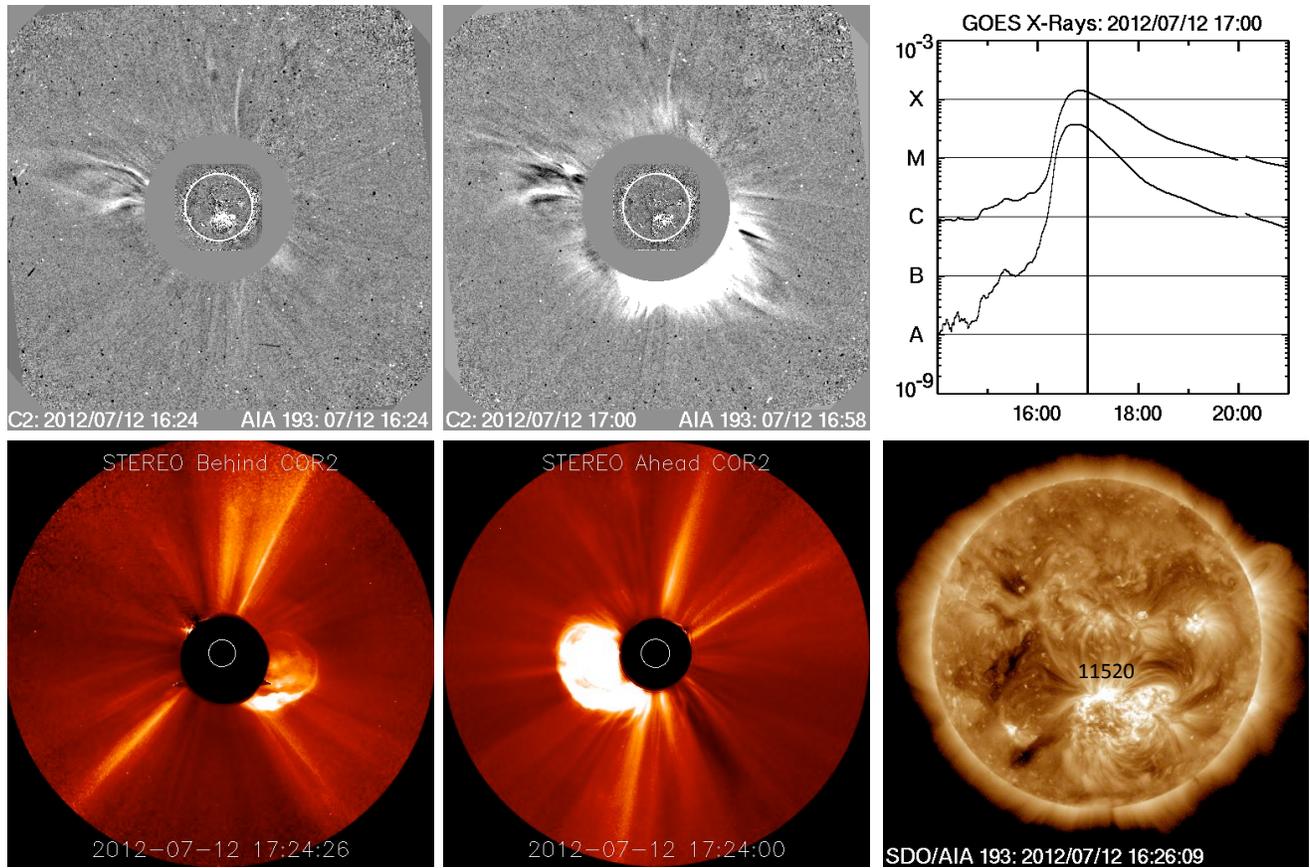

**Figure 1.** Images illustrating how we verify that a CME comes from an AR or group of ARs. The top two images are LASCO C2 images, the first one during the rise of the source AR flare, and the second one when the CME was clearly visible outside the C2 occulting disk. A corresponding GOES X-ray plot is shown in top right. STEREO-A (center) and B (left) images in the bottom row verify that the CME is Earthward directed. Last panel is an image taken from the AIA 193 Å movie, verifying that the position and NOAA number of the AR responsible for the CME are correct.

This careful manual selection procedure left a sample of 252 CMEs, with known flaring source ARs. The sample was further reduced by the requirement that there was available a definitive HMI AR Patch (HARP) vector magnetogram that covered the source AR, which was taken within 12 hours of the CME flare, and which had its magnetic flux centroid (defined below) within 45 heliocentric degrees of disk center. We also required that the source AR (1) be the only NOAA AR in the HARP tile, (2) the values of parameters are mostly (≥ 90%) from the AR and only negligibly (≤ 10%) from the other parts of the tile, or (3) if many ARs exist in the HARP, they are closely merged together and can be treated as one AR. This left a sample of 189 CMEs that we finally kept for our study.

We use the HARP vector magnetograms, which have better azimuthal disambiguation than the Space-Weather HMI AR Patches [SHARPs *Bobra et al.*, 2014]. From the HARP vector magnetograms, we measured the magnetic parameters described in the next section. The magnetograms have a pixel size of $0.5''$ and a cadence of 12 minutes. We prefer HARP over SHARP because our purpose in this paper is to look for any relationship between magnetic parameters and speed of CMEs that might lead to improvements in ongoing/future forecasting tools e.g., MAG4 [*Falconer et al.*, 2014], in contrast to the aim of devising a near real time tool for forecasting the speed of CMEs.

Each HARP was deprojected to disk center, i.e., LOS and transverse vector components were transformed to vertical and horizontal vector components and resampled to square pixels. Noise from transverse field and foreshortening is prohibitive when ARs are far from disk center. Therefore, we limit our sample to HARPs within 45 heliocentric degrees. Falconer et al (2015, in preparation) lays out the center-to-limb increase in deprojection errors in detail and shows that the parameters studied here have acceptably small projection errors out to 45 heliocentric degrees.

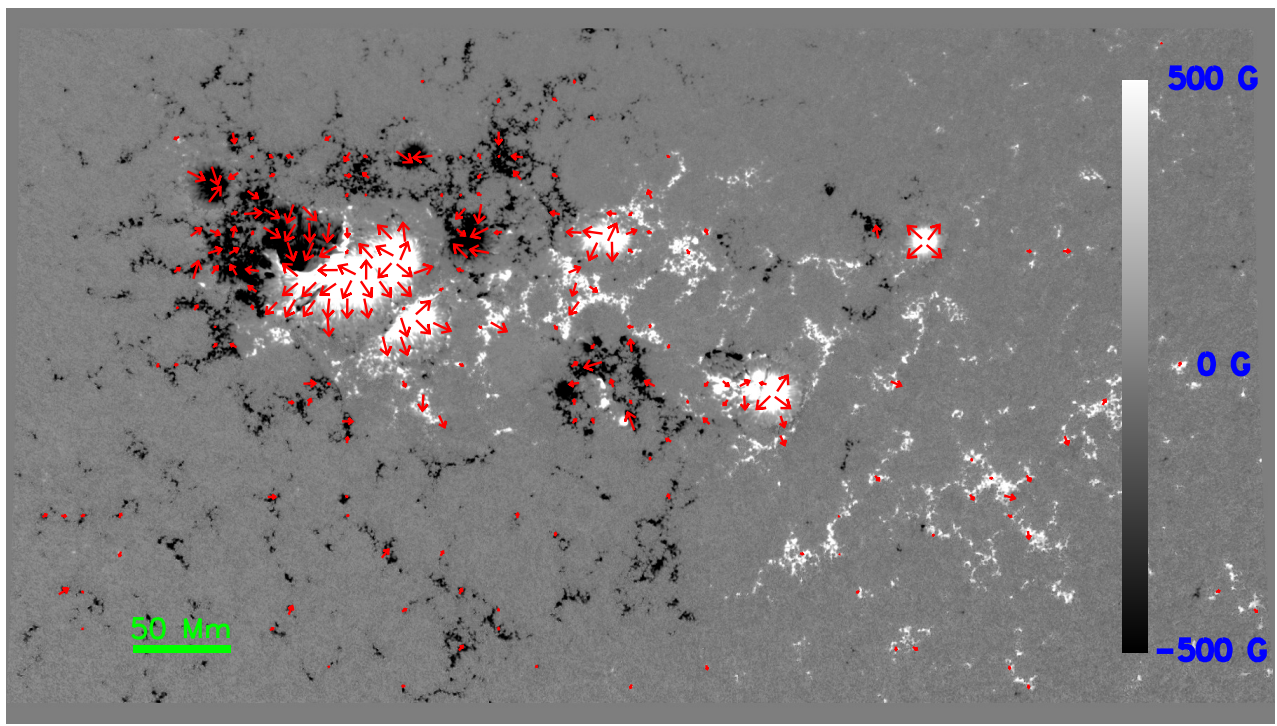

**Figure 2.** An example de-projected HARP-tile vector magnetogram containing NOAA AR 11520, which produced the X-class flare and CME shown in Figure 1. The size and direction of red vectors, overplotted on the grey-scaled vertical-field magnetogram, show the magnitude and direction of the horizontal field. The longest/shortest vector is for 500/100 G field strength.

In Figure 2, we display an example of a deprojected vector magnetogram tile. We have reduced noise in the measured parameters by using only pixels where the field components are above certain threshold values (see next Section).

The CME speeds have been obtained from the online LASCO/CME catalog [*Gopalswamy et al.*, 2009]. We use the speeds that are obtained from the linear fits to the height-time plot of the CME front in the plane of the sky [*Yashiro et al.*, 2004; *Gopalswamy et al.*, 2009]. The uncertainty in this measurement of the speed is less than 10% [*Yashiro et al.*, 2004; *Gopalswamy et al.*, 2012, Yashiro, 2015, private communication]. This is apart from the basic deficiency of using 2-D images in contrast to using speeds calculated from 3-D reconstruction of the CMEs [e.g., *Joshi and Srivastava*, 2011; *Mishra and Srivastava*, 2013]. The difference between the true speed and the measured plane-of-sky speed is found to be more for the CMEs with smaller widths [e.g., *Gopalswamy et al.*, 2012, Yashiro, 2015, private communication]. As mentioned before, our CMEs are wider than 30°, therefore not exposed to larger error from projection on the plane-of-sky. We have included both halo and non-halo CMEs to investigate the general correspondence between the speed of CMEs and the magnetic non-potentiality of the source ARs. Because the CME speeds used are plane-of-sky speeds, they are smaller than the true speeds of the CME fronts. It is worth mentioning here that the estimation of true speeds of Earth-directed CMEs is difficult. In a case study, by using stereoscopic observations, *Gopalswamy et al.* [2012] found the plane-of-sky speed measured by LASCO to be smaller by only 7.6% and 3.4% than the plane-of-sky speeds measured by STEREO-A and STEREO-B, respectively.

## 3. AR Magnetic Parameters Studied

### 3.1. AR Size Parameters

We use two AR size parameters. Both are integrals of all pixels that have absolute vertical magnetic field strength $B_z$ greater than 100 G. The first is the total magnetic area A,

$$A = \int dA, \quad (1)$$

and the second is the total magnetic flux $\Phi$,

$$\Phi = \int B_z dA. \quad (2)$$

### 3.2. Length of Strong-Field Neutral Line

The strong-field neutral-line length of an AR is defined by

$$L_S = \int dl, \quad (3)$$

where the integral is over all intervals of neutral lines in which the horizontal component of the potential field is greater than 150 G, and the interval separates opposite polarities of at least 20 G field strength [*Falconer et al.*, 2008]. These neutral-line intervals are used for the two other neutral-line-length parameters, which are free-energy proxies, described in Section 3.4.

To avoid magnetic parameters being dominated by noise, in this study we measure only ARs that are what we define to be strong-field ARs. Our definition follows *Falconer et al.* [2009]: a strong-field AR is one for which the ratio of $L_S$ to the square root of the magnetic area A is greater than 0.7.

### 3.3. Global Twist Parameters

**Global Alpha ($\alpha_g$):** The magnetic twist parameter $\alpha$ measures the vertical gradient of magnetic twist (radians of twist per unit length of height) in each pixel of a deprojected AR vector magnetogram [see Appendix A of *Tiwari et al.*, 2009b]; see also *Leka and Skumanich* [1999]. A global value of $\alpha$ can be calculated using the following formula [e.g., *Tiwari et al.*, 2009b]:

$$\alpha_g = \frac{\sum(\frac{\partial B_y}{\partial x} - \frac{\partial B_x}{\partial y})B_z}{\sum B_z^2} \quad (4)$$

We use this direct way of obtaining global $\alpha$ because the singularities at neutral line are automatically avoided in this method by using the second moment of minimization. Only pixels with absolute $B_z$ greater than 100 G are included in $\alpha_g$.

**Signed Shear Angle (SSA):** Motivated by the presence of oppositely directed twists at small-scales in sunspot penumbrae, *Tiwari et al.* [2009a] proposed SSA, which measures magnetic twist in ARs irrespective of their force-free nature [*Tiwari et al.*, 2009a] and shape [*Venkatakrishnan and Tiwari*, 2009]. It can be computed for each pixel of the deprojected vector magnetograms from the following formula:

$$SSA = \tan^{-1}(\frac{B_{yo}B_{xp} - B_{yp}B_{xo}}{B_{xo}B_{xp} + B_{yo}B_{yp}}) \quad (5)$$

where $B_{xo}$, $B_{yo}$ and $B_{xp}$, $B_{yp}$ are the observed and potential horizontal components of sunspot magnetic fields, respectively. The potential field is calculated from the vertical magnetic field using the method of *Alissandrakis* [1981]. Only pixels with absolute $B_z$ greater than 100 G are used.

Spatially averaged SSA (SASSA) and the median of SSA (MSSA) are each a global magnetic twist parameter of an AR. The difference between the two is the following: noisy pixels contribute directly to SASSA, whereas they are least weighted for MSSA. Therefore we have treated MSSA as a third global twist parameter here.

The SASSA and MSSA are both signed parameters; however in the present study only magnitude is taken into account.

### 3.4. AR Free-energy Proxies

**Gradient-weighted Neutral Line Length (WL$_{SG}$):** introduced by *Falconer et al.* [2008], this neutral line length measure is defined as

$$WL_{SG} = \int |\nabla Bz| dl \quad (6)$$

where $|\nabla Bz|$ is the horizontal gradient of the vertical magnetic field. The integral is computed for all neutral-line intervals that separate opposite polarities of at least moderate field strength of 20 G and have horizontal potential field greater than 150 G. Please note that these cutoff values are based on those taken by *Falconer et al.* [2008] for MDI data, and smaller numbers can be chosen for the HMI data. However, to be on the safe side, we have kept the same cutoff values in our present study.

**Shear-weighted Neutral Line Length ($WL_{SS}$):** also introduced in *Falconer et al.* [2008], this parameter is given by

$$WL_{SS} = \int |\phi - \phi_p| dl \quad (7)$$

where $\phi$ is the azimuth angle of the observed horizontal magnetic field, and $\phi_p$ is the azimuth angle of the potential horizontal magnetic field computed from the vertical magnetic field. The two free-energy proxies ($WL_{SG}$ and $WL_{SS}$) are strongly correlated, and are being explored in another work to determine which parameter is better for flare prediction.

**Schrijver's-R:** *Schrijver* [2007] developed a free-energy proxy that measures the amount of flux near neutral line pixels. To obtain Schriver's-R, which we denote as $R_{Schr}$, first a neutral-line pixel map is determined. This is done by determining all pixels are near a neutral line and that have positive or negative flux greater than 150 G. This step identifies strong-gradient neutral lines. This strong-gradient neutral-line-pixel map is then convolved with a 15 Mm Gaussian (as defined in *Schrijver* [2007] for MDI resolution). $R_{Schr}$ is the unsigned flux in that area divided by that area, giving $R_{Schr}$ a unit of G (Gauss). See *Schrijver* [2007], for more detail.

**Net Current:** The vertical current density $J_z$ can be measured from a deprojected vector magnetogram using the following formula:

$$J_z = \frac{1}{\mu_0}\left(\frac{\partial B_y}{\partial x} - \frac{\partial B_x}{\partial y}\right). \quad (8)$$

An integration of $J_z$ over all strong-field pixels ($|B_z| > 100$ G or $B_h > 200$ G) of an AR provides the net current for that AR. Following *Ravindra et al.* [2011], we use the sign convention that positive current flows upward in positive polarity regions, and downward in negative polarity regions, with negative current having the opposite flow. To obtain the net current $I_z$, the net current in the positive polarity pixels, and the net current in the negative polarity pixels (see Figure 2), are added and divided by 2. This is the net current and not the total current since ARs can easily have, in one part of a polarity domain, positive current flowing and in other parts have negative current [e.g., *Ravindra et al.*, 2011].

In addition to the above free-energy proxies, by multiplying by AR magnetic flux $\Phi$, we converted the twist parameters $\alpha_g$, SASSA and MSSA to the AR free-energy proxies $\alpha_g \times \Phi$, SASSA$\times\Phi$, MSSA$\times\Phi$. This is meaningful because an AR with large twist but little total flux plausibly does not have enough free-energy to produce fast CMEs, but an AR with the same large twist and large-enough flux plausibly does have enough free-energy to produce fast CMEs.

## 4. Results and Discussion

In Figure 3, total unsigned magnetic flux, three twist parameters and six free-energy proxies (two of which are combinations of twist and flux) of ARs are plotted against the plane-of-sky speed of CMEs emanating from the ARs. For all the ten plots in Figure 3 most data points fill a triangle portion of the phase space. Dashed red lines, drawn by eye following *Venkatakrishnan and Ravindra* [2003] (see dashed line in their Figure 3), outline the triangle area in each panel to roughly trace the upper bound of the speeds of CMEs.

Three important features in the plots that determine how well the speed of the fastest CMEs arising from an AR can be predicted are 1. the y-intercept of the red dashed line, 2. number of outliers above the line, and 3. how far above in y-direction the outliers are from the line. By the y-intercept of the red dashed line in each plot of Figure 3, we mean the y value at the point of intersection of the dashed line with the y-axis of that plot (the left side of the box).

The triangular shape of the clouds of plotted points shows that the ARs with large non-potentiality and large total flux produce both fast and slow CMEs, whereas ARs with the lower non-potentiality and less flux produce only slower CMEs. This behavior is similar to the behavior that the most non-potential ARs capable of producing large X-class flares also produce many smaller M-, and C- class flares, and ARs with relatively small non-potentiality rarely if ever produce larger flares [e.g., *Tiwari et al.*, 2010].

For most plots there are a few outliers that are above the line. For two plots, $\alpha_g$ and $\alpha_g \times \Phi$, the upper limit line is not strongly violated and the upper-limit CME velocity for the smallest of these two parameters (y-intercept) is $\sim$300 km s$^{-1}$. The number of outliers and their distance in y-direction from the line is least for these two parameters. The magnetic flux plot shows a limit line of similar low y-intercept, but has more outliers, which are relatively farther in y-direction above the limit line.

For the two better performing parameters, $\alpha_g$ and $\alpha_g \times \Phi$, the red dashed lines in Figure 3 give:

$$v = 10^{(2.48+0.42\times log_{10}(\alpha_g/2\times 10^{-10}))} \text{ km s}^{-1}, \text{ and} \quad (9)$$

$$v = 10^{(2.48+0.31\times log_{10}(\alpha_g\times\Phi/2\times 10^{12}))} \text{ km s}^{-1}, \text{ respectively.} \quad (10)$$

Because the lines are drawn by eye, they are not the only ones that could be drawn. By drawing different limit lines however we find no improvements in the predictive capabilities of the other eight parameters. For example, a less steep slope on FEP4 (Figure 3) can reduce the number of outliers but it also increases the y-intercept of the limit line significantly. Similarly the steepness of line can be increased in plots of the other

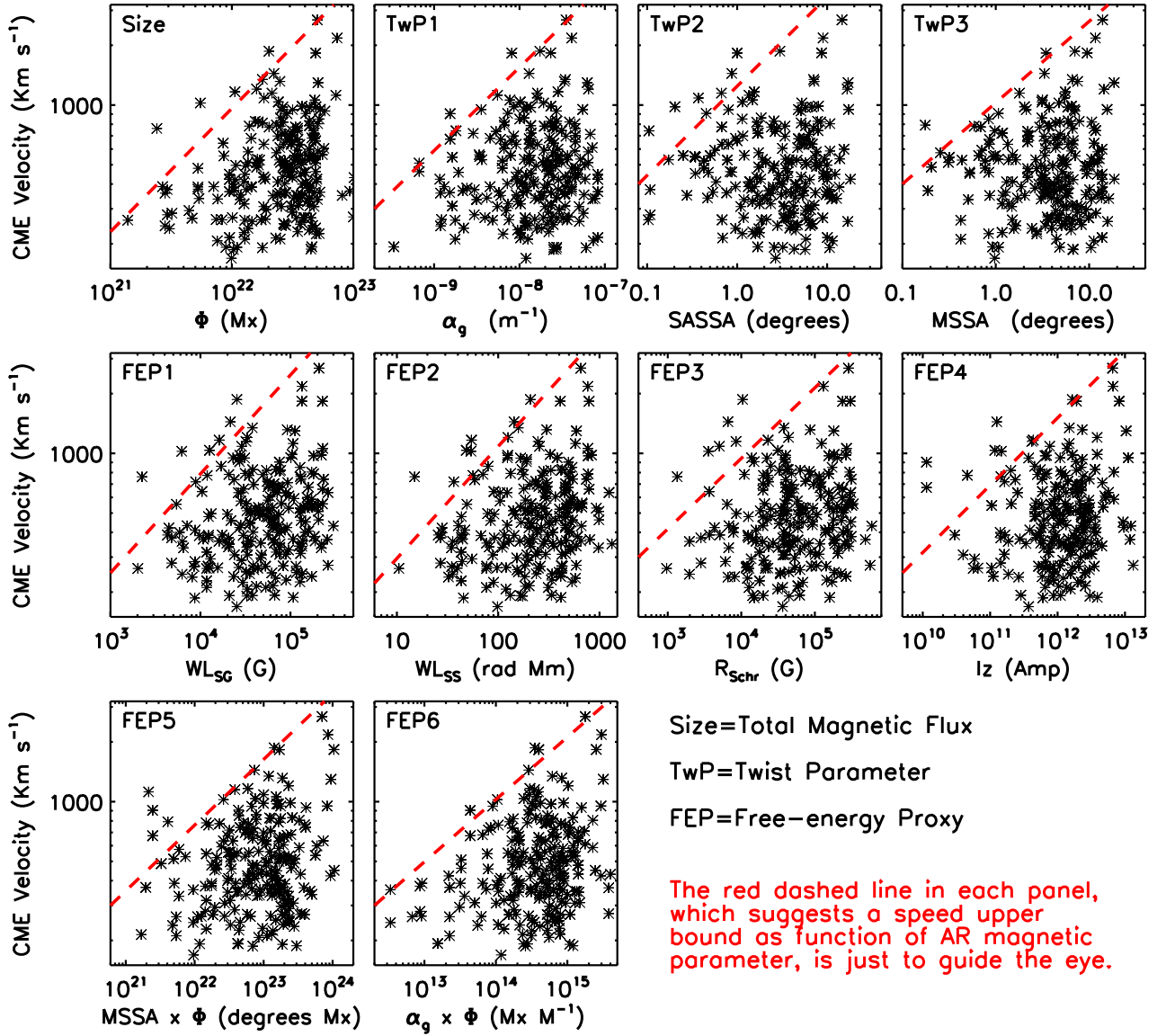

**Figure 3.** Scatter plots, in logarithmic scales on both x- and y-axes, of CME speed vs 10 different magnetic parameters of the source ARs. The first plot is for the AR's magnetic size (total magnetic flux $\Phi$), the next three plots are for whole-AR magnetic twist parameters ($\alpha_g$, SASSA, MSSA), and the last six plots are for AR free-energy proxies (WL$_{SG}$, WL$_{SS}$, R$_{Schr}$, I$_z$, MSSA$\times\Phi$, $\alpha_g \times \Phi$).

seven of the eight parameters but that increases the number of outliers.

The $\alpha_g$ limit line has a lower y-intercept and all of outliers are as close or closer to the limit line in y-direction, than for other two twist parameters, of which MSSA does better than SASSA. The fact that the $\alpha_g$ is weighted by strong magnetic field values and not affected by singularities at neutral lines [*Tiwari et al.*, 2009b, a] might be responsible for its superior behavior over the other two twist parameters. The MSSA does better than the SASSA probably because while taking median, a few noisy pixels with extremely high values of SSA are suppressed whereas they contribute more to SASSA.

The neutral-length free-energy proxies do not directly include the full area of the ARs [*Falconer et al.*, 2008], and display some outliers. The same is true for R$_{Schr}$. The limit line for net current shows a number of outliers. The net current varies from zero to nonzero values [*Venkatakrishnan and Tiwari*, 2009; *Ravindra et al.*, 2011; *Vemareddy et al.*, 2015] in different phases of AR's lifetime. The evolution of net current could possibly explain why this free-energy proxy is not the best for predicting the upper speed limit of CMEs in a statistical sense.

The curent solar cycle has been weak and we do not have many CMEs faster than 1000 kms$^{-1}$ in our sample. By extending the sample as more data becomes available in the LASCO/CME catalog we will determine if this result is robust, or if the speed limit edge for $\alpha_g$ and $\alpha_g \times \Phi$ in Figure 3 becomes less sharp.

From the results of *Venkatakrishnan and Ravindra* [2003] and *Liu* [2007], we expect the free-energy proxies to better determine the upper speed limit of CMEs that an AR can produce than twist parameters do. The fact that the twist parameter $\alpha_g$ displays nearly similar limit line as its corresponding free-energy proxy $\alpha_g \times \Phi$, is surprising and remains to be explained.

In line with our observations, numerical simulations also suggest that the same ARs can produce both fast and slow CMEs, with the most complex ones producing the fastest CMEs [see e.g., *Török and Kliem*, 2007]. The origin of slow CMEs from ARs with large non-potentiality can probably be explained by the fact that oftentimes only a part of AR takes part in the eruption leading to a CME and the full non-potentiality of the AR does not drive those CMEs. However, identifying the exact part of the ARs that produces a flare/CME is not an easy task due to their complex magnetic structuring.

In the present analysis, we have used free-energy proxies instead of computing free magnetic energy itself that requires vector magnetograms measured in the force-free field above the photosphere, which are not available owing to instrumental limitations, and also to the lack of reliable STOKE's profiles inversion codes for NLTE atmospheres. This limitation can be partially overcome by reliable non-linear force-free field modeling [*Wiegelmann and Sakurai*, 2012; *Wiegelmann et al.*, 2014] based on the photospheric vector field measurements of ARs, which are not entirely force-free on the AR photosphere [*Metcalf et al.*, 1995; *Tiwari*, 2012] but can be preprocessed to make them force-free under certain circumstances. Future studies should make use of such techniques to improve the accuracy of the prediction of the upper speed limit of the CMEs that an AR can produce.

## 5. Conclusions

In this Letter, we investigated the correspondence between the speed of CMEs and non-potentiality of their source ARs by using a total of 189 CMEs.

Plane-of-sky speed of CMEs were taken from the SOHO/LASCO CME Catalog. In addition to total unsigned magnetic flux, various magnetic twist parameters and free energy proxies of the source ARs were measured to gauge their non-potentiality. To measure these parameters, HARP vector magnetograms from HMI were used after deprojection onto the solar disk center.

We find a general trend among all parameters that the ARs with larger non-potentiality and total magnetic flux can produce both fast and slow CMEs, whereas the ARs with smaller non-potentiality and flux can only produce slower CMEs. There are exceptions present for all of the parameters. Out of all the parameters studied, $\alpha_g$ and $\alpha_g \times \Phi$ show the best triangular pattern with least outliers, and lowest y-intercept of the limit line, thus conveying their better performance over the other parameters for predicting the upper limit of the speed of CMEs that an AR can produce.

Since fast CMEs tend to be a greater threat for severe space weather than slower ones, knowing that an AR can not produce a fast CME would be a useful forecast. Thus, our results can be incorporated in near real time forecasting tools e.g., MAG4 [*Falconer et al.*, 2014]. Expanding the data set of CMEs having measured speeds and measureable source ARs in future will improve statistics and confirm or modify our results.

**Acknowledgments.** SKT would like to thank Tibor Török and Yang Liu for useful discussion on this work during the AGU-2014 meeting. We acknowledge Phyllis Whittlesey and Malte Broese, students from Joint Space Weather Summer Camp 2014 (sponsored by the University of Alabama in Huntsville and its Center for Space Plasma and Aeronomic Research, The University of Rostock and its Leibniz-Institute of Atmospheric Physics, and the German Aerospace Center (DLR)), for initially identifying some of the CMEs and their source ARs. SKT is supported by an appointment to the NASA Postdoctoral Program at the NASA MSFC, administered by ORAU through a contract with NASA. RLM and ARW are supported by funding from the LWS TRT Program of the Heliophysics Division of NASAs SMD. Support for MAG4 development comes from NASA's Game Changing Development Program, and Johnson Space Center's Space Radiation Analysis Group (SRAG). Use of SOHO LASCO CME catalog, and data from AIA and HMI (SDO), STEREO, SOHO is sincerely acknowledged. The CME catalog is generated and maintained at the CDAW Data Center by NASA and The Catholic University of America in cooperation with the NRL. SOHO is a project of international cooperation between ESA and NASA.

Corresponding author: SANJIV K. TIWARI, NASA Marshall Space Flight Center, ZP 13, Huntsville, AL 35812, USA (sanjiv.k.tiwari@nasa.gov)